\documentclass[aps,prl,twocolumn,groupedaddress]{revtex4}
\usepackage{graphicx}
\usepackage{amsmath,amsfonts,amssymb}
\usepackage{color}
\usepackage{float}
\begin{document}

\title{Frozen up Dilaton and the GUT/Planck Mass Ratio}

\author{Aharon Davidson}
\email{davidson@bgu.ac.il}
\author{Tomer Ygael}
\email{tomeryg@post.bgu.ac.il}
\affiliation{Physics Department, Ben-Gurion University
of the Negev, Beer-Sheva 84105, Israel}
\date{\today}

\begin{abstract}
	By treating modulus and phase on equal footing, as prescribed
	by Dirac, local scale invariance can consistently accompany any
	Brans-Dicke $\omega$-theory.
	We show that in the presence of a soft scale symmetry breaking
	term, the classical solution, if it exists, cannot be anything else but
	general relativistic.
	The dilaton modulus gets frozen up by the Weyl-Proca vector
	field, thereby constituting a gravitational quasi-Higgs mechanism.
	Assigning all  grand unified scalars as dilatons, they enjoy Weyl
	universality, and upon symmetry breaking, the Planck (mass)$^2$
	becomes the sum of all their individual (VEV)$^2$s.
	The emerging GUT/Planck (mass)$^2$ ratio is thus
	$\sim \omega g_{GUT}^2/4\pi$.
\end{abstract}

\pacs{04.50.Kd, 14.70.Pw, 12.10.Dm}

\maketitle

\noindent \textbf{Critical local Weyl invariance}\smallskip

The Brans-Dicke theory \cite{BD} is described by the action
\begin{equation}
	{\cal I}_{BD}=-\int d^4 x\sqrt{-g}\left(\phi^2 {\cal R}
	+4\omega g^{\mu\nu}\phi_{;\mu}\phi_{;\nu}
	\right) ~.
	\label{BD}
\end{equation}
The theory, characterized by a dimensionless parameter $\omega$,
the coefficient of the dilaton kinetic term, is invariant under the
combined \emph{global} scaling transformation
\begin{equation}
	g_{\mu\nu}(x) \rightarrow e^{-2\chi}g_{\mu\nu}(x) ~, \quad
	\phi(x)  \rightarrow e^{\chi}\phi(x)  ~.
\end{equation}
Consistent with the latter global symmetry is the quartic scalar
potential term $V(\phi)=\lambda \phi^4$.
As is well known, it is only the critical case $\omega=-\frac{3}{2}$
which further enjoys the full \emph{local} scale symmetry
\begin{equation}
	g_{\mu\nu}(x) \rightarrow e^{-2\chi(x)}g_{\mu\nu}(x) ~, \quad
	\phi(x)  \rightarrow e^{\chi(x)}\phi(x)  ~.
	\label{chix}
\end{equation}
In the 'unitary' gauge, often called Einstein gauge, defined by fixing
$\phi(x)=v$, for some arbitrary constant $v$, the theory resembles
general relativity chracterized by an arbitrary
Planck mass $M^2_{Pl}=16\pi v^2$, and furthermore accompanied by a matching 
cosmological constant $\Lambda_E=\frac{1}{2}\lambda v^2$.
This by itself, however, does not make Einstein theory of gravity a
gauge-fixed version of the critical Bran-Dicke theory.

\medskip\noindent \textbf{Non-critical local Weyl invariance}\smallskip

Recalling the profound success of the standard electro/nuclear
theory, a local scale symmetry is most welcome and currently quite
popular \cite{local1,local2}.
The relative minus sign between the gravitational and the kinetic
scalar terms, a characteristic feature of the critical Brans-Dicke
theory, appears to be problematic on ghost related grounds.
However, as we were guided by Dirac \cite {Dirac}, local scale
invariance can be extended to accompany any Brans-Dicke
$\omega$-theory, including in particular the $\omega>0$
(no ghost) branch.
The Dirac prescription reads
\begin{equation}
	{\cal I}_{D}=-\int d^4 x~\sqrt{-g}\left(\phi^2 {\cal R}{^*}
	+4\omega g^{\mu\nu}\phi_{*\mu}\phi_{*\nu}
	\right) ~,
	\label{DBD}
\end{equation}
instructing us to replace the various tensors involved
by their (starred) co-tensor substitutes.
The procedure requires the presence of the Weyl vector field
$\kappa_{\mu}$, subject to the familiar transformation law
\begin{equation}
	\kappa_{\mu}(x) \rightarrow \kappa_{\mu}(x)
	-\chi (x)_{;\mu} ~.
	\label{Schi}
\end{equation}
An optional universal coupling constant has been momentarily
absorbed within $\kappa_{\mu}$ redefinition (to be justified later
on universality grounds).

The Ricci scalar replacement ${\cal R}^*$ takes the explicit form
\begin{equation}
	{\cal R}^*={\cal R}-6g^{\mu\nu}\kappa_{\mu;\nu}
	+6g^{\mu\nu}\kappa_{\mu}\kappa_{\nu} ~.
\end{equation}
Under scale transformations it behaves as a co-scalar of power -2,
that is ${\cal R}^* \rightarrow e^{2\chi(x)}{\cal R}^*$.
While the dilaton field $\phi$ is by construction a co-scalar of
power -1, its covariant derivative $\phi_{;\mu}$ is not a co-vector
at all.
It is only the co-covariant Weyl derivative \cite{Wder}
\begin{equation}
	\phi_{*\mu}=\phi_{;\mu}+ \kappa_{\mu}\phi ~,
	\label{phistar}
\end{equation}
which constitutes a co-vector of power -1, thereby making the
kinetic term replacement $g^{\mu\nu}\phi_{*\mu}\phi_{*\nu}$
a legitimate co-scalar of power -4.
The Weyl co-covariant derivative conceptually differs from the
Stueckelberg \cite{Stueckelberg} covariant derivative
$\phi_{;\mu}+m \kappa_{\mu}$, but
is in full analogy with the Maxwell covariant derivative
$\phi_{;\mu}+i e A_{\mu}\phi$ of an electrically charged (and
hence necessarily complex) scalar field.
The imaginary electromagnetic coupling constant $ie$
has been forcefully traded for a real (currently absorbed as noted
earlier) coupling constant.

A mandatory ingredient is a kinetic term for the Weyl vector field.
Truly, it is not directly required on plain local scale symmetry
grounds, but in its absence $\kappa_{\mu}$ would have stayed
non-dynamical in nature.
The transformation law eq.(\ref{Schi}) dictates the exact Maxwell
structure, with the corresponding anti-symmetric differential
2-form given by
\begin{equation}
	X_{\mu\nu}=\kappa_{\mu;\nu}-\kappa_{\nu;\mu} ~.
\end{equation}
Altogether, up to the total derivative
$6(\phi^2 \kappa ^{\mu})_{;\mu}$, and a full
re-arrangement of the various terms floating around, the
non-critical (arbitrary $\omega$) local Weyl invariant theory
can be described in a somewhat more familiar language by
the action
\begin{equation}
	\begin{array}{ccc}
	&  \displaystyle{{\cal I}=-\int d^4x\sqrt{-g}
	 \left[\phi^2\left({\cal R}-6s
	g^{\mu\nu}\kappa_{\mu}\kappa_{\nu}\right)
	+\right.} &   \vspace{4pt}\\
	&  \left.+~4\omega g^{\mu\nu}
	D_{\mu}\phi D_{\nu}\phi+\lambda\phi^4
	+\frac{1}{4}g^{\mu\nu}g^{\lambda\sigma}
	X_{\mu\lambda}X_{\nu\sigma}\right]~, &   
	\end{array}
	\label{newBD}
\end{equation}
where we have used the shorthand notation
\begin{equation}
	s=\frac{3+2\omega}{2\omega} \neq 1~.
\end{equation}
The latter action eq.(\ref{newBD}) looks deceptively
conventional, so a word of caution is necessary.
Note that
\begin{equation}
	D_{\mu}\phi=\phi_{;\mu}+s \kappa_{\mu}\phi 
	\label{D}
\end{equation}
is in fact a \emph{fake} co-covariant derivative, and should
not be confused with the \emph{genuine} co-covariant
derivative eq.(\ref{phistar}).

In the Einstein gauge $\phi(x)=v$ (with $v$ still being
an arbitrary constant at this stage),
the theory resembles a particular Einstein-Proca \cite{Proca}
theory accompanied as before by a cosmological
constant $\Lambda_E=\frac{1}{2}\lambda v^2$
(note that the corresponding Proca/Planck mass ratio is
$v$-independent).
However, this by itself does not make Einstein-Proca theory
of gravity a gauge-fixed version of the non-critical Bran-Dicke
theory.
A gravitational Higgs-like mechanism capable of singling out the
'unitary' Einstein gauge on physical (local scale) symmetry
breaking grounds is in order.
In this paper, however, only a pseudo-Higgs mechanism is offered.

\medskip\noindent \textbf{A fake scale symmetry?}\smallskip

The notion of a fake symmetry has been coined by Jackiw
and Pi \cite{JackiwPi} to address a situation where the
conserved Noether/Weyl current vanishes identically.
Their assertion was that in certain cases the corresponding
Weyl symmetry does not actually have any dynamical role.
The critical Brans-Dicke theory, as well as some of its
currently proposed derivative models \cite{local2}, fall into
such a category.
Following the Jackiw-Pi analysis, we now calculate the
Noether/Weyl current stemming from the action
eq.(\ref{newBD}).
The result is non-trivial, owing to the presence of $\kappa_{\mu}$,
thereby implying in our case a genuine local Weyl symmetry.

First, without using the Euler-Lagrange equations, we perform
the variation with respect to the combined
symmetry transformations eqs.(\ref{chix},\ref{Schi}), and find
\begin{equation}
	\delta {\cal L}=L^{\mu}_{~;\mu}\sqrt{-g} ~,
	\quad L^{\mu}=-6\phi^2g^{\mu\nu}\chi_{;\nu} ~.
	\label{L}
\end{equation}
Utilizing a previous Jackiw-Pi calculation, the Weyl vector field
$\kappa_{\mu}$ and its antisymmetric derivative $X_{\mu\nu}$
simply do not enter at this stage.
Next, however, we do invoke the equations of motion, and
following the Noether procedure, conventionally use them to
eliminate
$\displaystyle{\frac{\partial {\cal L}}{\partial \phi}}$,
$\displaystyle{\frac{\partial {\cal L}}{\partial \kappa_{\mu}}}$, and
$\displaystyle{\frac{\partial {\cal L}}{\partial g_{\mu\nu}}}$
from the variation.
Doing so, we arrive at an alternate divergence formula for
$\delta {\cal L}$ which can be written in the form
\begin{equation}
	\delta {\cal L}=(L^{\mu}+J^{\mu})_{;\mu}\sqrt{-g} ~.
	\label{LJ}
\end{equation}
Equating eqs.(\ref{L},\ref{LJ}), the Weyl/Noether  conservation
law makes its appearance
\begin{equation}
	J^{\mu}_{~;\mu}=0 ~.
\end{equation}
But this time, contrary to the fake symmetry case, the classically
conserved symmetry current $J^{\mu}$ does not vanish.
To be specific, it is explicitly given by
\begin{equation}
	J^{\mu}=8\omega \phi^2 \kappa^{\mu} \chi
	+sX^{\mu\nu}\chi_{;\nu} ~,
\end{equation}
emphasizing the role played by the Weyl gauge field.
Associated with the Lagrangian eq.(\ref{newBD}) is thus a
genuine local scale symmetry.

\medskip\noindent \textbf{Frozen up dilaton}\smallskip

A spontaneously scale symmetry breaking mechanism in
four (generically in more than two) dimensions is still at
large.
The emergence of the Planck mass scale within the
framework of a theory which does not tolerate the
introduction of any dimensional parameter at the level
of the bare Lagrangian is quite challenging \cite{Shaposhnikov}. 
For a recent attempt, based on a Coleman-Weinberg
like mechanism in a framework similar to ours, see
Ref.(\cite{Ohanian}).
With this in mind, we leave the kinetic part of the Lagrangian
absolutely intact, and thus fully scale symmetric, and supplement
the potential part by a soft scale symmetry breaking piece.
By 'soft' we mean

\noindent $\bullet$ Terms whose coefficients have a positive
power of mass, 

\noindent $\bullet$ Terms whose transformation law do not
involve derivatives of the gauge function $\chi(x)$.

\noindent 
In particular, while a scalar field mass term is welcome, a
vector field mass term will not do.
Following 'tHooft \cite{tHooft10}, adding a
non-conformal part such as a scalar field mass term does
not have any effect on the dangerously divergent term in
the effective action.
The more so in the context of this paper, adding a scalar
field mass term should be regarded merely a technical tool
primarily designed to single out the Einstein gauge.
We will show that the corresponding classical solution, if it
exists, cannot be anything else but general relativistic.

We thus return to the action eq.(\ref{newBD}), and would like
to trade the strictly quartic potential $\lambda \phi^4$ for a
more general potential of the type
$V(\phi)=\lambda \phi^4 + p \phi^2+q$.
On pedagogical grounds, however, to appreciate the fact that
our results are generic, and are not that sensitive to the exact
structure of the potential, we keep momentarily working with
a general $V(\phi)$. 
Associated with the non-critical Weyl invariant Lagrangian
contaminated by a general scalar potential $V(\phi)$ are the
following field equations, corresponding to variations with
respect to $\phi,\kappa_{\mu},g_{\mu\nu}$, respectively:
\begin{subequations}
\begin{eqnarray}
	&& 4\omega g^{\mu\nu}\phi_{;\mu\nu}=
	\textstyle{\frac{1}{2}} V^{\prime}(\phi)+
	\nonumber \\ [2pt]
	&&~+{\phi\cal R}+4\omega s \phi
	g^{\mu\nu}(\kappa_{\mu}\kappa_{\nu}
	-\kappa_{\mu;\nu}) ~, \label{phi} \\[8pt]
	&&X^{~\nu}_{\mu;\nu}=4s\omega\left(
	\phi^2_{~;\mu} +2 \kappa_{\mu}\phi^2
	\right) ~, \label{S}\\ [8pt]
	&&\phi^2\left({\cal R}_{\mu\nu}
	-\textstyle{\frac{1}{2}} g_{\mu\nu}{\cal R}\right)=
	\nonumber\\ [2pt]
	&&~=-\phi^2_{~;\mu\nu}
	+g_{\mu\nu}g^{\alpha\beta}\phi^2_{;\alpha\beta}
	+\textstyle{\frac{1}{2}}  g_{\mu\nu}V(\phi)-
	\nonumber\\ [2pt]
	&&~-\gamma_{\mu\nu}+\textstyle{\frac{1}{2}} g_{\mu\nu}
	g^{\alpha\beta}\gamma_{\alpha\beta}- 
	\nonumber\\ [3pt]
	&&~-\textstyle{\frac{1}{2}} X_{\mu}^{~\lambda}X_{\nu\lambda}
	+\textstyle{\frac{1}{8}} g_{\mu\nu}
	X^{\alpha\beta}X_{\alpha\beta} ~,
	\label{g}
\end{eqnarray}
\end{subequations}
where we have used the notation
\begin{equation}
	\gamma_{\mu\nu}=-6 s
	\phi^2 \kappa_{\mu}\kappa_{\nu}
	+4\omega D_{\mu}\phi D_{\nu}\phi ~.
\end{equation}
We can now trace eq.(\ref{g}) and subsequently substitute
the Ricci scalar ${\cal R}$ into eq.(\ref{phi}).
Re-organizing the various terms, we arrive at a generalized
Klein-Gordon equation for $\phi^2$, namely
\begin{equation}
	g^{\mu\nu}\left(\phi^2_{~;\mu}
	+2 \kappa_{\mu}\phi^2
	\right)_{;\nu}
	=\frac{\partial W_{eff}(\phi^2)}{\partial\phi^2}~.
	\label{eff}
\end{equation}
The effective potential $W_{eff}(\phi^2)$ which governs the
$\phi^2$-evolution, defined by means of
\begin{equation}
	\frac{\partial W_{eff}(\phi^2)}{\partial\phi^2}
	=\frac{1}{3+2\omega}
	\left(\frac{1}{2}\phi V^{\prime}(\phi)-2V(\phi)\right) ~,
\end{equation}
is known to play a central role \cite{Veff} in scalar-tensor theories.
One may verify that, owing to its conformal nature, the quartic term
$\lambda \phi^4$ in $V(\phi)$ does not contribute to $W_{eff}(\phi^2)$.

By no coincidence, the same current which sources eq.(\ref{S}), 
and whose divergence must therefore vanish identically (thereby
defining a superpotential), is exactly the current whose divergence
we meet again on the l.h.s. of eq.(\ref{eff}).
In turn, on self-consistency grounds, a classical solution can
exist only provided
\begin{equation}
	\boxed{\frac{1}{2}\phi V^{\prime}(\phi)-2V(\phi)=0}
	\label{Veq}
\end{equation}
We remark in passing that the chain of arguments leading to
eq.(\ref{Veq}) breaks down in the absence of the $\kappa_{\mu}$
gauge field.
In which case, the theory resembles Zee's broken symmetric
theory of gravity \cite{Zee}.
The above equation can also be written in the form
$V^{\prime}_E(\phi)=0$, where
\begin{equation}
	V_E(\phi)=\phi^{-4}V(\phi)
\end{equation}
stands for the associated Einstein frame scalar potential.
The crucial point now is that eq.(\ref{Veq}) constitutes an
\emph{algebraic} (rather than \emph{differential}) equation
for $\phi$. 
Its solutions, if exist, are constants rather than functions of $x$.
Other classical solutions, associated with (say) $\phi(x)$ evolving
along the scalar potential (e.g. oscillations around the VEV), simply
cannot exist.
Depending on the number of solutions, our discussion trifurcates:

\noindent (1) No solutions - The classical equations of motion are
self-inconsistent.
Not too many examples of this sort appear in the literature.

\noindent (2) A single solution - The classical configuration comes
then with a frozen dilaton field
\begin{equation}
	\phi_{cl}(x)=v ~,
\end{equation}
and thus, is exclusively general relativistic. 
The Planck mass being identified as $M^2_{Pl}=16\pi v^2$,
and the accompanying cosmological constant is given by
\begin{equation}
	\Lambda=\frac{1}{2v^2}V(v)
	=\frac{1}{8v}V^{\prime}(v) ~.
\end{equation}

\noindent (3) Multiple solutions - With eq.(\ref{Veq}) strictly
enforced, one cannot classically tell in this case a minimum from
a maximum.
In turn, each solution comes with its own Planck scale, a situation
to be avoided or hopefully resolved at the quantum mechanical level.

Resembling the phrase 'eaten up' borrowed from Higgs terminology,
the dilaton degree of freedom has been totally 'frozen up', thereby
converting the massless Weyl gauge field into a massive Proca field
\begin{equation}
	m^2_{\kappa}=-6s v^2+4\omega s^2 v^2
	=4\omega s v^2 ~,
\end{equation}
leaving the physical spectrum scalar particle free.
Altogether, starting from an explicit (soft) scale symmetry
breaking, we have eventually encountered a gravitational
quasi-Higgs mechanism.
We note that both spherically symmetric \cite{ProcaBH} as well
as cosmological \cite{ProcaFLRW} solutions of the  Einstein-Proca
theory have already been studied in the literature. 

Where does the consistency requirement eq.(\ref{Veq}) actually
come from?
The points to notice are that (i) By construction,  the $\kappa_{\mu}$
field equation does not directly 'know' about $V(\phi)$.
In particular, its associated conserved current stays independent
of $V$ and $V^{\prime}$.
Hence, the $\kappa_{\mu}$ equation still captures the full local
scale invariance of the kinetic part of the Lagrangian.
(ii) Treating eq.(\ref{Veq}) as a differential equation for $V(\phi)$
results in the conformal $V(\phi)=\lambda \phi^4$.
But once a different potential enters the game, affecting only the
$\phi,g_{\mu\nu}$-equations of motion, the local scale symmetry
identity turns an algebraic constraint which is solely respected by
the Einstein gauge..

Clearly, the local structure of $V(\phi)$ is irrelevant at this level.
It is only a global feature, namely the discrete spectrum of the
$\displaystyle{\frac{\partial W_{eff}(\phi^2)}{\partial\phi^2}}$ roots,
which actually matters.
Insisting on soft scale symmetry breaking, we restrict ourselves
to the class of bi-quadratic polynomials.
It is practical to parametrize the potential as follows
\begin{equation}
	V(\phi)=\lambda \phi^4+(2\Lambda-\lambda v^2)
	(2\phi^2-v^2) ~.
\end{equation}
Such a potential has the further advantage that its effective potential
companion (up to a non-physical additive constant)
\begin{equation}
	W_{eff}(\phi^2)=\frac{\lambda v^2-2\Lambda}{3+2\omega}
	(\phi^2-v^2)^2 ~,
\end{equation}
admits a single extremum (as a function of $\phi^2$).
Note that the positivity of $W$ is correlated with the negativity
of the $\phi^2$ mass term added.
While classically, as explained, we cannot tell a minimum from
a maximum, quantum mechanical stability would
require $\Lambda <\Lambda_E= \frac{1}{2}\lambda v^2$ for
a ghost free (positive) $\omega$.
This includes the special $\Lambda=0$ case.

\medskip\noindent
\textbf{CP-violating Weyl-Maxwell mixing}\smallskip
 
By construction, the local scale invariant non-critical Brans-Dicke
theory can easily accommodate a complex dilaton field.
The action eq.(\ref{newBD}) is simply traded for
\begin{equation}
	\begin{array}{ccc}
	&  {\cal I}=-\int d^4x\sqrt{-g}
	 \left[\phi^{\dagger}\phi
	 \left({\cal R}-6s g^{\mu\nu}\kappa_{\mu}\kappa_{\nu}\right)
	 +\right. & \vspace{4pt}\\
	&  \left.+4\omega g^{\mu\nu}
	(D_{\mu}\phi)^{\dagger} (D_{\nu}\phi)
	+\lambda(\phi^{\dagger}\phi)^2
	+\frac{1}{4}g^{\mu\nu}g^{\lambda\sigma}
	X_{\mu\lambda}X_{\nu\sigma}\right]~, &   
	\end{array}
	\label{complexBD}
\end{equation}
leaving the door open for the incorporation of Abelian 
and non-Abelian gauge fields.
Using the notation
\begin{equation}
	\phi(x)=\rho(x) e^{i\theta(x)} ~,
\end{equation}
with eq.(\ref{D}) becoming now
\begin{equation}
	D_{\mu}\phi=e^{i \theta}\left(\rho_{;\mu }
	+s \rho \kappa_{\mu }+i \rho\theta _{;\mu }\right)~.
\end{equation}
In turn, the action eq.(\ref{complexBD}) takes now the exact
form of eq.(\ref{newBD}), with $\rho$ replacing $\phi$ of course,
to which the generalized kinetic term $4\omega \rho^2g^{\mu\nu }
\theta _{;\mu }\theta _{;\nu }\sqrt{-g}$ is added.
Note that the realization of the Einstein gauge $\rho(x)=v$ is
achieved without restricting $\theta(x)$ whatsoever, leaving the
latter to play the role of a free massless scalar field in the Einstein
frame.
Still, using the notion of a Goldstone boson, while quite tempting,
is unjustified here since no global symmetry has actually been
spontaneously violated.

The subsequent incorporation of a $U(1)$ gauge interaction
is achieved naturally and flawlessly.
However, from an obvious reason (soon to be clarified), it should
be emphasized from the outset that it cannot be electromagnetism
we are talking about.
Associated with a $U(1)$-charged dilaton is now the unified
co-covariant derivative
$\phi_{\star\mu}=\phi_{;\mu}+\kappa_{\mu}\phi+ie A_{\mu}\phi$.
By the same token, the modified fake co-covariant derivative splits
into
\begin{equation}
	D_{\mu}(\rho e^{i\theta})=e^{i\theta}\left[\rho_{;\mu}
	+s \rho \kappa_{\mu}
	+i \rho (\theta_{;\mu}+eA_{\mu})\right]~,  
\end{equation}
with the non-Abelian generalization being straight forwards
(and relevant for a variant class of grand unified theories).
A key element in allowing for this construction is the fact that
$A_{\mu}$ constitutes a co-vector of power zero.
In turn, as was shown by Dirac \cite{Dirac}, its co-covariant
derivative $A_{\mu \star \nu}$ differs from the covariant derivative
$A_{\mu;\nu}$ only by a symmetric term, namely
\begin{equation}
	A_{\mu \star \nu}=A_{\mu;\nu}
	+\kappa_{\mu}A_{\nu}+\kappa_{\nu}A_{\mu}
	-g_{\mu\nu} \kappa^{\alpha}A_{\alpha} ~,
\end{equation}
so that also
$F_{\mu\nu} \equiv A_{\mu \star \nu}-A_{\nu \star \mu}$
stays power zero.

The Maxwell kinetic term, on the other hand, can be
accompanied by a novel mixed kinetic term
\begin{equation}
	\frac{1}{4}g^{\mu\nu}g^{\lambda\sigma}
	X_{\mu\lambda}X_{\nu\sigma}
	+\frac{1}{4}g^{\mu\nu}g^{\lambda\sigma}
	F_{\mu\lambda}F_{\nu\sigma}
	+\frac{\xi}{2}g^{\mu\nu}g^{\lambda\sigma}
	F_{\mu\lambda}X_{\nu\sigma} ~,
	\label{mixing}
\end{equation}
parametrized by a dimensionless coefficient $\xi$,  which
cannot be ruled out solely on local symmetry grounds.
On group theoretical grounds, a non-Abelian analogue simply
cannot exist.
Note that a $U(1)\otimes U(1)$ kinetic mixing has already
been studied by Holdom \cite{Holdom} as a mechanism for
shifting electromagnetic charges by a calculable amount.

Soft scale symmetry breaking, the advocated mechanism for
singling out general relativity at the classical level, is governed
now by the $U(1)$-invariant scalar potential
\begin{equation}
	V(\phi)=\lambda (\phi^{\dagger}\phi)^2
	+(2\Lambda-\lambda v^2)
	(2\phi^{\dagger}\phi-v^2) ~.
\end{equation}
A closer inspection reveals that once the dilaton modulus
$\rho(x)$ gets frozen up and the Goldstone boson $\theta(x)$
eaten up, one encounters a diagonal (mass)$^2$ matrix for the
two vector fields involved, namely
\begin{equation}
	m^2_{\kappa}=2(3+2\omega)v^2  ~,
	\quad m^2_A=4\omega e^2 v^2  ~.
\end{equation}
Notably, unlike the conventional Higgs mechanism, the
physical spectrum does not contain a massive free scalar
particle, to be regarded a fingerprint of the pseudo-Higgs
mechanism.
The non-diagonal kinetic mixing eq.(\ref{mixing}) gives
rise to the Holdom effect.
The equations of motion involve two conserved currents
\begin{eqnarray}
	&& X^{~\nu}_{\mu;\nu}+\xi F^{~\nu}_{\mu;\nu}
	=8\omega s\left[\frac{1}{2} (\phi^{\dagger}\phi)_{;\mu}
	+\kappa_{\mu}\phi^{\dagger}\phi \right]
	~,\quad\\
	&& F^{~\nu}_{\mu;\nu}+\xi X^{~\nu}_{\mu;\nu}
	=4\omega e g^{\mu\nu}
	\left[-i \phi^{\dagger} \overleftrightarrow{\nabla_{\mu}}
	\phi+2eA_{\mu}\phi^{\dagger}\phi \right] ~.
\end{eqnarray}
In particular, a residual non-vanishing $U(1)$ source current,
proportional to $(1-\xi^2)^{-1}\xi v^2 \kappa_{\mu}$, survives
the $e \rightarrow 0$ limit.
Moreover, contrary to the Holdom mixing, the Weyl-Maxwell
mixing is in fact CP-violating.
This comes about when noticing \cite{CP}, as indicated
by the structure of the conserved currents, that under a
CP transformation
\begin{equation}
	A_{\mu}\rightarrow -A_{\mu}~, \quad
	\kappa_{\mu}\rightarrow +\kappa_{\mu}~.
\end{equation}

The masses $m^2_{\kappa ,A}$ are proportional to a common
VEV, the one which sets the Planck scale $M^2_{Pl}=16\pi v^2$
in the present theory.
This is the reason why $A_{\mu}$ cannot represent here
electromagnetism.
From the same reason, the standard model Higgs doublet
cannot serve as a dilaton (like in the Higgs inflation scenario
\cite{inflation}).
It is more likely that a grand unified theory (GUT)
is involved.
In which case, $m^2_{GUT}$ and $M^2_{Pl}$ share a
common origin, and hence acquire the one and the same
mass scale.
Up to some potentially large group theoretical factor
$\sim 10^{2-3}$, associated with the dimensions and
multiplicity of the scalar field representations involved, the
typical (mass)$^2$ ratio should be
\begin{equation}
	\boxed{\frac{m^2_{GUT}}{M^2_{Pl}}
	\propto\omega\frac{g_{GUT}^2}{4\pi} }
	\label{ratio}
\end{equation}
where $g_{GUT}$ stands for the coupling constant of
the grand unifying group.

\medskip\noindent \textbf{Weyl universality and the Planck scale}
\smallskip

A grand unified theory generically introduces a variety of scalar
fields.
And once local scale symmetry joins the game, the question is
whether such a unified theory can tolerate the coexistence of
several kinds of dilatons $\phi_i$, differing from each other not
only by their grand unified representation $\underline{r_i}$ but
also by their scaling powers.
The answer of course is negative, and the reason is quite obvious.
We may have several scalars at our disposal, but just one
underlying metric to govern the dynamics of the spacetime they
live in.
To be specific, owing to the identical structure of their kinetic
terms (exhibiting a single $g^{\mu\nu}$), all minimally coupled
scalar fields must constitute co-scalars of order -1. 
By a similar token, all massless fermions involved constitute
co-spinors of order -3/2, transforming according to
$\psi \rightarrow e^{\frac{3}{2}\chi} \psi$.
Their kinetic terms
\begin{equation}
	\int d^4 x \sqrt{-g}
	i\bar{\psi}\gamma^{\mu} (x)
	[\partial_{\mu}+\Gamma_{\mu}(x)]\psi ~,
\end{equation}
with $\Gamma_{\mu}(x)$ denoting the Levi-Civita spin connection,
have the further advantage that they are automatically conformally
invariant.
In turn, unlike the universal minimal coupling of the scalar fields,
\emph{fermions simply do not couple to the Weyl gauge vector
field $\kappa_{\mu}$}.

We now attempt to go one step further and suggest a variant
grand unified theory where all scalar fields are in fact dilatons.
Following the Dirac prescription, their individual local scale
symmetric contributions to the kinetic term in the Lagrangian
sum up into
\begin{equation}
	-\int d^4 x \sqrt{-g}\left[{\cal R}^{\star}
	\sum_i \phi_i^{\dagger}\phi_i
	+4 g^{\mu\nu}\sum_i \omega_i
	\phi_{i\star\mu}^{\dagger}\phi_{i\star\nu}
	\right] ~,
\end{equation}
where $\phi_{i\star\mu}=(\nabla_{\mu}
+I \kappa_{\mu}+i g T^k_i A_{k\mu})\phi_i$.
Note that local scale symmetry can tolerate
$\omega_i \neq \omega_j$ for $i\neq j$.
Such an arbitrariness in the Weyl sector remind us, in a remote
way, of a similar arbitrariness which characterizes the Yukawa
sector.
While the latter formula is just a straight forwards generalization
of eq.(\ref{DBD}), its overall message is pleasing and is by no
means conventional:
The Planck (mass)$^2$ which governs the general relativistic
coupling of matter to geometry is nothing but the sum over all
individual (VEV)$^2$s which have been invoked to give mass to
the variety of particles in the first place.
To be more specific,
\begin{equation}
	\boxed{M_{Pl}^2
	=16\pi \sum_i v_i^{\dagger}v_i}
	\label{Planck}
\end{equation}
This formula is clearly in accord with the GUT/Planck mass ratio
eq.(\ref{ratio}), and by being sensitive to the underlying group
theoretical structure (expressed via the sum), can hopefully be
used to tell one such grand unified theory from the other.
A particular $SO(10)$ grand unified model incorporating the
latter idea is currently in the make.

\medskip\noindent \textbf{Epilogue}\smallskip

Local \emph{scale} symmetry and its celebrated relative local
\emph{phase} symmetry should be treated on equal theoretical
footing.
As far as the Brans-Dicke theory is concerned, the fact that
local scale symmetry is not restricted any more to the critical
case $\omega=-\frac{3}{2}$, but can virtually accompany any
$\omega$ variant, opens the door for revisiting a full range
of imaginative theoretical ideas.
A leading such idea views general relativity as a spontaneously
generated theory of gravity.
Its spontaneous scale symmetry breakdown mechanism is however
still unknown, at least at the practical level.
In this respect, the soft (explicit, but without affecting the divergent
term in the effective action) scale symmetry breaking hereby invoked,
 while being just a poor man's alternative, it is nonetheless
 sophisticated enough to provide us with a novel gravitational
 quasi-Higgs mechanism where the dilaton modulus gets frozen
 up (thereby defining the Newton constant) by the Weyl-Proca vector
 field.
The latter acquires the Planck mass scale and becomes the
physical fingerprint of the emergent general relativity.

\acknowledgments
{One of us (A.D.) cordially thank Prof. Ray Volkas for his generous
hospitality at the University of Melbourne. Fruitful discussions
with him and with David Wakeham are acknowledged.
We also thank Eduardo Guendelman for his insight concerning
CP-violation.}

\end{document}